\documentclass[10pt,journal,compsoc]{IEEEtran}
\usepackage{array}
\usepackage{enumerate}
\usepackage{amsfonts}
\usepackage{multirow}
\usepackage{amsmath}
\usepackage{amssymb}
\usepackage{graphicx}
\usepackage{url}
\usepackage{bbm}
\usepackage{placeins}

\RequirePackage{fix-cm}
\usepackage[linesnumbered,ruled,vlined]{algorithm2e}
\usepackage{tabulary}



\newtheorem{prop}{Proposition}

\newtheorem{note}{Note}

\SetAlgorithmName{Protocol}{Protocol}{}
\begin{document}

\title{Circuit-Free General-Purpose 
Multi-Party Computation via Co-Utile Unlinkable Outsourcing}

\author{Josep Domingo-Ferrer,~\IEEEmembership{Fellow, IEEE,} and Jes\'us Manj\'on%
\thanks{The authors are with Universitat Rovira i Virgili,
Department of Computer Engineering and Mathematics,
CYBERCAT-Center for Cybersecurity Research of Catalonia,
UNESCO Chair in Data Privacy,
Av. Pa\"{\i}sos Catalans 26, E-43007 Tarragona, Catalonia.
e-mail \{josep.domingo, jesus.manjon\}@urv.cat}}

\markboth{Vol.~?, No.~?, Month~YYYY}%
{J. Domingo-Ferrer \MakeLowercase{\textit{et al.}}: General Multi-Party
Computation via Co-Utility} 

\IEEEtitleabstractindextext{%
\begin{abstract}
Multiparty computation (MPC) consists in
several parties engaging in joint computation
in such a way
that each party's input and output remain private
to that party. Whereas MPC protocols
for specific computations have existed since the 1980s,
only recently general-purpose compilers have been developed
to allow MPC on arbitrary functions.
Yet, using today's MPC compilers requires substantial
programming effort and skill on the user's side, among 
other things because nearly 
all compilers translate the code of the computation
into a Boolean or arithmetic circuit.
In particular, the circuit 
representation requires unrolling loops and recursive
calls, which forces programmers to (often manually) 
define loop bounds
and hardly use recursion.
We present an approach allowing MPC on an arbitrary
computation expressed as ordinary code with all functionalities
that does not need to be translated
into a circuit.
Our notion of input and output privacy is predicated on unlinkability.
Our method leverages co-utile computation outsourcing using
anonymous channels via decentralized reputation,
makes a minimalistic use of cryptography
and does not require participants to be honest-but-curious: it
works as long as participants are rational
(self-interested), which may include rationally malicious peers
(who become attackers if this is advantageous to them).
 We present example applications, including e-voting.
Our empirical work shows that reputation captures well the behavior
of peers and ensures that parties with high reputation obtain
correct results.
\end{abstract}

}

\maketitle

\IEEEpeerreviewmaketitle

\section{Introduction}
\label{introduction}

\IEEEPARstart{M}{ultiparty} computation (MPC) consists in
several parties engaging in joint computation
in such a way
that each party's input and output remain private
to that party.
As phrased in~\cite{hastings},
MPC can be viewed as a cryptographic method for 
providing the functionality of a trusted party ---whose
role would be to accept private inputs from a set
of participants, compute a function
and return the outputs to the corresponding participants---
without the need for mutual trust among participants.

MPC can be used to solve a variety of problems that 
require using the data of participants without compromising
their privacy. Example applications include secure statistical analysis,
financial oversight, electronic voting, secure machine learning, 
auctions, biomedical computations (in particular 
those involving highly sensitive data, such as genetic information), etc. 
See~\cite{evans,lindell,hastings} for more details and
related references on MPC applications.

The most central security properties required in MPC are as follows:
\begin{itemize}
\item {\em Privacy}. No party should learn anything more than
its prescribed output. In particular, nothing should be learnt
about the other parties' inputs except what can be derived from
the output itself.
\item {\em Correctness}. Each party should receive its prescribed
output and this output should be correct.
\end{itemize} 

There are different security models to characterize
the assumptions on the adversarial behavior of parties in MPC.
In the {\em honest-but-curious} model 
(also called {\em semi-honest}), parties are assumed to 
correctly follow the protocol specification, but they may
try to learn other parties' inputs or outputs. 
In the {\em malicious} model, parties can arbitrarily 
deviate from the protocol
specification. As pointed out in~\cite{evans},
the honest-but-curious model is very weak and it underestimates
the power of realistic adversaries in most scenarios. The
malicious model, in contrast, is very strong but only 
a small subset of MPC protocols in the literature 
can cope with it~\cite{furukawa}. 
In this paper, we focus on a third security model, namely
the {\em rational} model, in which parties are self-interested: 
they only deviate from the protocol specification if doing so
is more advantageous to them than following the protocol.
Thus, this model includes 
rationally malicious parties, who may become attackers if this is advantageous
to them. Hence, the rational model is stronger than honest-but-curious
but weaker than malicious (which allows irrationally malicious behavior).

Since the 1980s, when the seminal MPC protocols~\cite{Y,GMW,BGW,CCD}
were proposed, and until very few years ago, 
practical MPC protocols existed only
for specific computations. 
The appearance of general-purpose compilers enabling MPC for 
arbitrary functions is a recent breakthrough; see~\cite{hastings} for 
a state of knowledge on such compilers.   
Most MPC compilers translate ordinary high-level programming code expressing
the computation to be performed into a Boolean or an arithmetic circuit.
Once at the circuit level, they use several cryptographic primitives, such
as secret sharing, oblivious transfer, garbled circuits and others 
to generate the MPC protocol for the target computation 
(see~\cite{evans} for descriptions of such primitives).

As noted in~\cite{hastings}, using state-of-the-art compilers
requires substantial programming effort and skill on 
the user's side, due {\em inter alia} to the inherent constraints
of the circuit representation.
In particular, loops and recursive calls in the original 
computation code must be unrolled, which forces programmers
to (often manually) define loop bounds and hardly use recursion.

In this paper we propose an MPC protocol that has the 
following distinguishing properties:
\begin{itemize}
\item It is general-purpose without making use of circuits; 
it can work for any computation expressed
as ordinary high-level programming code, no matter the depth
of its loops, its use of recursion or its complexity.
\item It does not rely on the cryptographic primitives usual in MPC;
it uses cryptography only to guarantee confidential 
and authenticated communications.
\item It assumes a peer-to-peer (P2P) community is available,
where each peer accumulates a reputation
in a decentralized and {\em co-utile} manner.
\item It relies on the notion of {\em co-utile} anonymous channel to guarantee
that inputs and outputs, although
visible to some peers, cannot be linked to their corresponding parties.
\item It delivers correct exact outputs as long as parties
are {\em rational}.
\end{itemize}
We illustrate our approach in two example applications, 
one of them being electronic voting.
Our empirical work shows that reputation captures well the behavior
of peers and ensures that parties with high reputation obtain
correct results.

Section~\ref{co-utility} gives some background on co-utility
and co-utile decentralized reputation.
Section~\ref{suite} introduces a co-utile protocol suite
for MPC.
Section~\ref{discoutil} discusses how the proposed protocol
suite achieves co-utility (and thus is rationally sustainable).
Section~\ref{privsec} shows how our protocols satisfy privacy
for the inputs and the outputs, and how they make
 correct computation  the best option for rational peers.
Section~\ref{applications} sketches illustrating applications.
Section~\ref{experiments} presents experimental results.
Finally, Section~\ref{conclusions} deals with conclusions 
and future research.

\section{Co-utility and co-utile decentralized reputation}
\label{co-utility}

A self-enforcing protocol
is \emph{co-utile}~\cite{coutility} if it results in mutually
beneficial collaboration between the participating agents.
More specifically, a protocol $\Pi$ is co-utile
if and only if {\em the three} following conditions hold:
(i) $\Pi$ is self-enforcing; (ii) the utility derived 
by each agent participating
in $\Pi$ is strictly greater than the utility the agent
would derive from not participating; and (iii) there is 
no alternative protocol $\Pi'$
giving greater utilities to all agents
and strictly greater utility to at least one agent.

The first condition ensures that if participants
engage in the protocol, they will not deviate.
The second condition is needed to ensure that engaging
in the protocol is attractive for everyone.
The third condition can be
rephrased in game-theoretic terms by saying that the
protocol is
a Pareto-optimal solution of the underlying game.

In~\cite{InfSciRep}, we gave a co-utile adaptation
of the EigenTrust decentralized reputation protocol~\cite{Kamvar03}.
In addition to co-utility, this protocol has 
the following attractive properties:
decentralization, pseudonymity of peers, low overhead, 
 proper management of new peers (to discourage whitewashing
bad reputations as new identities) and resistance to reputation
tampering (such as self-promotion or slandering).

The protocol consists in calculating a global reputation
for each peer $P$ based on aggregating the local opinions of the
peers that have interacted with $P$.
If we represent the local opinions by a matrix $(\ell_{ij})$ 
where $\ell_{ij}$ contains the opinion of peer $P_i$ on peer $P_j$,
the distributed calculation mechanism computes global reputation
values that approximate the left
principal eigenvector of this matrix.
We will use a 
co-utile decentralized reputation system which can be viewed
as a simplified adaptation of that of~\cite{InfSciRep}.

\section{A co-utile framework
for multi-party computation}
\label{suite}

\subsection{Players and security model}
\label{players}

The players in our framework are peers in a P2P network as follows: 
\begin{itemize}
\item {\em Clients} are the parties that agree on a joint 
computation to be conducted, and then provide private inputs 
 to it and obtain  private outputs from it. 
The number of clients must be at least 4: with only 2 clients, if a 
client sees an input or an output that is not hers, 
she knows it corresponds to the other client; with only 3 clients,
unequivocal inferences are still possible, as justified in our
privacy analysis of Section~\ref{secpriv}. 
\item {\em Workers} perform computations for clients.
\item {\em Forwardees} receive
messages from clients and forward them to other forwardees
or submit them to workers.
\item {\em Accountability managers}
are peers that manage the reputations of other peers in the network.
Clients, workers and forwardees all have accountability managers.
Each peer $P$ is assigned $M$ accountability managers that 
are (pseudo)randomly determined by hashing the 
peer's pseudonym $P$. In this way, $P$ cannot choose her 
 accountability managers, which makes the latter more likely
to perform their duty impartially and therefore honestly.
\end{itemize}

In our framework, the inputs provided by clients 
and the outputs they obtain cannot be unequivocally linked 
 to them, even though such inputs
and outputs may be seen by some other peers.
Our privacy guarantee is predicated
on unlinkability rather than on confidentiality.
{\em If inputs or outputs are such that 
it is not acceptable to disclose them even unlinkably 
or such that their very values or formats make them linkable to certain
clients, 
then our framework cannot be used.}

{\em Clients are publicly known to each other} (they are neither   
anonymous nor pseudonymous), because in general a peer wants to know
with whom she is engaging in joint computation.
Forwardees, workers 
and accountability managers are pseudonymous.
A client also uses a pseudonym to interact with forwardees,
workers and accountability managers; clients do not know
each others' pseudonyms. 
A peer's pseudonym $P_i$ and her public identity $ID_i$ are
related as
$P_i = H(ID_i, nonce_i)$,
where $H(\cdot)$ is a one-way hash function and $nonce_i$
is a random number only known to peer $ID_i$; in this way,
the pseudonym $P_i$ does not leak the underlying real
identity $ID_i$, but the link between both can be proven
if necessary by revealing $nonce_i$.
During a protocol execution, {\em any peer
can play any of the above four roles} (client, worker,
forwardee, accountability manager).

We assume that {\em peers are rational}:
 given appropriate incentives they will honestly fulfill
their roles in the protocol, but they may be curious
to learn the inputs or outputs of specific clients.
We will design protocols so 
that these rational peers find no incentive to deviate.
However, there may be a minority of  
malicious peers that are {\em irrational}, that is,
who are ready to deviate from the protocols even 
if doing so places them in a worse position.

Although we will show how to incentivize rational peers
to adhere to our protocols, we cannot guarantee
that {\em clients} will provide truthful inputs for any arbitrary
multi-party computation; this can only be guaranteed
if clients are assumed honest or honest-but-curious.
For rational clients, input truthfulness depends on the specific
MPC to be carried out.
In some MPCs, a rational client may 
find incentives to provide a false input, {\em e.g.} if 
she knows her false input cannot be detected by the others 
but it allows her to be the only client that learns the correct output.
However, in other MPCs the rational behavior is to provide correct
inputs. Take for example the millionaires' problem~\cite{Y}: 
if millionaire $A$
inputs an amount higher than her actual fortune and the output is
that $A$ is richer than the other millionaire $B$, then $A$ does
not learn whether she is really richer or poorer than $B$; similarly, if $A$
inputs an amount lower than her actual fortune and the output is
that $A$ is poorer than $B$, then $A$ does not learn whether
she is really poorer or richer than $B$.

The main aim of a rational client is to obtain a
correct output of the joint
computation and to keep her inputs and outputs confidential from the
other clients and the other pseudonymous peers. As we will show 
in the next sections, a client needs to have a high reputation
to fulfill the previous aim. 
The incentive for a rational worker or a rational forwardee to cooperate 
is to increase her reputation in order to be able to become a 
successful client. 
Finally, the incentive for a rational accountability manager
is to preserve fairness in the community. More details
on the rational behavior of all peers are given in the next sections.

In some applications reputation alone may be deemed an insufficient incentive,
 for example because working to build up a high reputation
entails substantial financial costs in equipment, bandwidth, electricity, 
etc. To mitigate this shortcoming, reputation can be 
periodically converted into payments. In fact,  
rational incentives to correct computation in the form of payments are proposed 
in~\cite{kupcu15}, where they are implemented by a central ``boss'',
and in~\cite{harz2019,nabi2020}, where they are embodied in smart contracts
in blockchains. However, unlike in our approach based on 
reputation only, incentives in
these proposals require
external payment infrastructures that may not (always) be available
or that may themselves be centralized or not rationally
sustainable. 

\subsection{Requirements}
\label{requirements}

At the end of the previous section we have mentioned reputation
as the ``currency'' that incentivizes peers.
In order for reputation to be effective, the following requirements
need to be fulfilled:
\begin{itemize}
\item {\em Reward.} If a worker correctly performs a certain
computation for a client, the worker's reputation must increase.
\item {\em Punishment.}  If a worker incorrectly performs 
a certain computation for a client, the worker's reputation
must decrease. Similarly, a client that does not follow
the protocols as prescribed must be punished with a reputation decrease.
\item {\em Probabilistic reward.} A peer acting as a forwardee
for inputs or outputs should be motivated by a nonzero probability of 
obtaining a reputation increase.
\item {\em Reputation utility.} Having high reputation
must be attractive for peers. 
Specifically, the higher the reputation of a client,
the easier it is for the client to  
 retrieve her outputs while preserving
her privacy. For a pseudonymous peer, a high reputation
is the way to become a successful client.
\item {\em Plausible deniability.} The        
input contributed by a client should
not be unequivocally linkable either to her public identity
or to her pseudonym. 
Similarly, the output
obtained by a client should 
not be unequivocally linkable either to her public identity
or to her pseudonym. 
Hence, a client should be able to plausibly deny that a certain 
input or output are hers.
Our input/output  privacy guarantee is thus
 stronger than mere pseudonymity.
\end{itemize}

\subsection{General-purpose MPC in the honest-but-curious model}

For the sake of clarity, we first present
a basic version of our protocol suite 
assuming that all peers are honest-but-curious.
In Protocol~\ref{pro1},
each client $P$ secretly selects a worker $P_w$ (who does
not know who $P$ is). 
Every worker receives the inputs of all
clients in an unlinkable way via the co-utile anonymous 
channel FWD-CH (Protocol~\ref{p2pan}).
Each client $P$ also sends to her $P_w$ in an unlinkable way via FWD-CH 
the computation
to be performed and an encryption key to be used by $P_w$ to return
the results to $P$ via the reverse anonymous channel 
REV-CH (Protocol~\ref{p2prev}). 
The idea of using an anonymous channel for MPC was first proposed
in~\cite{isha06}; the novelty here is that we present an anonymous
channel that does not depend on any 
central authority and is rationally sustainable.

\begin{algorithm}[]
Clients $ID_1, \ldots, ID_m$ among the $n$ peers (where $m, n$ are public
 and $4 \leq m \leq n$) know each 
other and agree to jointly perform
the computation
$(O_1, O_2, \ldots, O_m) = C(I_1, I_2, \ldots, I_m)$, 
where input $I_i$ and output $O_i$ must stay private to 
$ID_i$\;
\For{$i=1$ \KwTo $m$ {\bf in parallel}}
{Client $ID_i$ uses her pseudonym $P_i$\;
$P_i$ prunes the code $C$ into the part $C_i$ that computes $O_i$,
{\em i.e.}
$O_i = C_i(I_1, I_2, \ldots, I_m)$;
\tcc{In case $C_i$ requires knowledge of which of the inputs is $P_i$'s,
$I_i$ is also embedded in $C_i$} 
$P_i$ randomly and secretly selects a peer $P_{w_i}$ among
the $n$ peers as a worker\;
\For{$l=1$ \KwTo $n$}{
$P_i$ calls $\mbox{FWD-CH}(P_i,I_i||nonce_i,nil,P_l)$;
\tcc{$P_i$ sends her private input $I_i$ to all peers $P_l$ via the
$\mbox{FWD-CH}(\cdot)$ anonymous channel, so that in particular $P_{w_i}$
gets it. $I_i$ is appended a random nonce to make it unique.} 
}
$P_i$ calls $O_i=\mbox{FWD-CH}(P_i,PK_{w_i}(K_i),C_i,P_{w_i})$.
\tcc{$P_i$ sends to $P_{w_i}$ a key $K_i$ encrypted under $P_{w_i}$'s public key and the computation $C_i$ (a peer's public key may be her pseudonym). 
Then $P_{w_i}$ will return $O_i$ symmetrically 
encrypted under $K_i$ via the reverse channel.}
}
\caption{{\sc Honest-but-curious general-purpose MPC}}
\label{pro1}
\end{algorithm}

\begin{algorithm}[]
\tcc{Unlinkability in FWD-CH is achieved by random hopping among peers
until a peer submits the message to the destination worker $P_d$.}
Parameter $p \in [0,1]$\;
\eIf{$P_s =  P_d$}{
\eIf{$comp =  nil$ AND $msg$ carries a new nonce}{ 
$P_d$ extracts the input $I$ in $msg$ and appends it to $Ilist$\;
\tcc{When $comp=nil$, $msg$ is an input $I$; 
$Ilist$ is initially empty and is appended all inputs with
different nonces received by $P_d$ in FWD-CH calls with $comp=nil$}
}
{ 
$P_d$ waits until $Ilist$ contains $m$ different inputs\;
$P_d$ performs computation $out=comp(Ilist)$\;
$P_d$ cleans $Ilist$\;
$P_d$ decrypts $K=SK_d(msg)$\;
\tcc{When $comp\neq nil$ and all inputs have been received, $P_d$ performs 
computation $comp$
on the received $Ilist$; also, $msg=PK_d(K)$, where the symmetric key 
$K$ is to be used
by $P_d$ to encrypt the output of $comp$ before returning it} 
$P_d$ calls $\mbox{REV-CH}(P_d,E_K(out),comp,P_{prev})$\; 
\tcc{REV-CH returns the encrypted output of $comp=C_i$ to 
originator $P_i$ without knowing who $P_i$ is; $P_{prev}$ 
is the pseudonym of the peer from whom $P_d$ received $comp$}
}
}
{
\leIf{$P_s$ is the originator of $msg$}
{$p_{forward}=1$}{$p_{forward}=p$}
\tcc{The originator always hops but other peers hop with prob. $p$}
\eIf{$\mbox{Bernoulli}(p_{forward})=1$}
{
$P_s$ randomly chooses another peer $P'$\;
$P_s$ sends $(msg,comp)$ to $P'$\;
$P'$ calls $\mbox{FWD-CH}(P',msg,comp,P_d)$\;
}
{
$P_s$ directly sends $(msg,comp)$ to $P_d$\;
$P_d$ calls $\mbox{FWD-CH}(P_d,msg,comp,P_d)$.
}
}
\caption{$\mbox{FWD-CH}(P_s, msg, comp, P_d)$}
\label{p2pan}
\end{algorithm}

\begin{algorithm}[]
\tcc{REV-CH backtracks the hopping path of FWD-CH up to the client}
$P_s$ sends $(E_K(out),comp)$ to $P_{prev}$\;
\eIf(\tcc*[f]{$P_{prev}$ is the client}){$P_{prev}$ knows the key $K$}{
$P_{prev}$ decrypts $out$;
\tcc*[f]{$out$ is $P_{prev}$'s private output $O_{prev}$}
}
{
Let $P_{prev2}$ be the peer from whom $P_{prev}$ 
received the FWD-CH message with $comp$\;
$P_{prev}$ calls $\mbox{REV-CH}(P_{prev},E_K(out), comp, P_{prev2})$.
\tcc*[f]{$P_{prev}$ backtracks to $P_{prev2}$}
}
\caption{$\mbox{REV-CH}(P_s,E_K(out), comp, P_{prev})$}
\label{p2prev}
\end{algorithm}


\subsection{General-purpose MPC in the rational model}

In the rational model peers may deviate from their prescribed
behavior if they are not properly incentivized. We will add two
mechanisms to cope with this problem:
(i) redundancy to detect wrong computation or wrong forwarding;
(ii) decentralized reputation to reward correct computation
and correct forwarding, and punish wrong behavior.
The reputations of all peers are {\em public}.

In the protocols in this section, we assume 
the public-key cryptosystem used to encrypt to peers 
is {\em probabilistic}, that is,
it uses randomness so that an observer cannot
determine whether two ciphertexts correspond to the same
cleartext.
Also, we assume all messages encrypted under
this public-key cryptosystem  
can be made of the same length, if necessary by padding
the cleartext, so that ciphertext length cannot be used by an observer
to guess information on the cleartext.
Additionally, we assume peers can digitally sign
messages.

Protocol~\ref{pro1rat} is the version of Protocol~\ref{pro1} augmented
with redundancy ($r$ different workers 
are used by each client) and decentralized reputation.
In Protocol~\ref{pro1rat} we use co-utile versions 
C-FWD-CH (Protocol~\ref{p2panrat})
and C-REV-CH (Protocol~\ref{p2prevrat}) 
of the previous FWD-CH and REV-CH auxiliary protocols.
For the rest and unless otherwise said, 
the notations are the same as in Protocol~\ref{pro1}.

\begin{algorithm}[]
Clients $ID_1, \ldots, ID_m$ among the $n$ peers 
(where $m, n$ are public and $4 \leq m \leq n$)
know each other and agree to jointly perform
the computation
$(O_1, O_2, \ldots, O_m) = C(I_1, I_2, \ldots, I_m)$,
where input $I_i$ and output $O_i$ must stay private to
$ID_i$\;
\For{$i=1$ \KwTo $m$ {\bf in parallel}}{
 Client $ID_i$ uses her pseudonym $P_i$\label{pseudo}\;
$P_i$ prunes the code $C$ into the part $C_i$ that computes $O_i$,
{\em i.e.}
 $O_i = C_i(I_1, I_2, \ldots, I_m)$\;
$P_i$ secretly selects as workers $r$ peers $P_{i_1},\ldots,P_{i_r}$ 
randomly chosen among the $\kappa_i > r$ peers with closest 
reputation to $g_i$\;
\For{$l=1$ \KwTo $n$}{
$P_i$ calls $\mbox{C-FWD-CH}(P_i,PK_{l}(I_i||nonce_i), PK_{l}(nil), P_l)$\label{inputsend}\tcc*[f]{$P_i$ sends her input to all peers, including
$P_i$'s workers. $PK_l$ is encryption under $P_l$'s public key.}\;
}
\For{$k=1$ \KwTo $r$}{ 
$P_i$ calls 
$O_{i,k}=\mbox{C-FWD-CH}(P_i,PK_{i_k}(K_i), PK_{i_k}(C_i),P_{i_k})$\label{paspas}\;
\ForAll{accountability managers $AM$ of $P_i$}{
\If(\tcc*[f]{Not rewarding the first forwardee is punished}){$P_i$ does not show to $AM$ a reward receipt from the first forwardee}{
$AM$ assigns local reputation $\ell_{AM,i}=0$ to $P_i$\label{1pas14};}
}
}
Take as $O_i$ the most frequent value in $\{O_{i,1},\ldots,O_{i,r}\}$\;
\For{$k=1$ \KwTo $r$}{
\eIf{$O_{i,k}=O_i$ AND $O_i \neq nil$}{
$P_i$ assigns $\ell_{i i_k}=1$;\label{reprew} \tcc*[f]{Majority result is 
rewarded}
}
{
$P_i$ assigns $\ell_{i i_k}=0$\label{reppun}
\tcc*[f]{Minority result is punished}}

}
}
Call Protocol~\ref{globrep} to update the public global reputation
$g_i$ of each peer $P_i$.
\caption{\sc Rational general-purpose MPC}
\label{pro1rat}
\end{algorithm}

\begin{note}[On parameter $\kappa_i$]
\label{notekappa}
Parameter $\kappa_i$ is secret to each client $P_i$. If $\kappa_i \gg r$,
then workers are selected from a large set, which makes it more difficult
for $P_i$'s workers to guess that their client is $P_i$.
On the other hand, too large a $\kappa_i$ is also risky for $P_i$,
because it can result in choosing workers with reputation much higher 
than $P_i$'s (who are likely to refuse working for $P_i$) or workers
with much lower reputation (who may be unreliable).
\end{note}

\begin{note}[On sharing workers]
A way to reduce the computation and communication of Protocol~\ref{pro1rat}
while still providing redundancy would be for clients to share
their workers, rather than each client choosing her own $r$ workers. 
For example, if each client proposed one worker, 
a pool of $m$ workers would be available to every client; to achieve
the same redundancy level, Protocol~\ref{pro1rat} requires using $m^2$
workers.
Yet sharing workers comes at a price. 
First, redundancy based on sharing workers means that the same computation $C$
must be run for all clients. 
Second, clients are forced to trust
workers that have been selected by other clients.
This is particularly bad if the reputations
of clients are very heterogeneous: high-reputation clients 
could find workers with higher reputation 
(and thus more reliable) than the workers 
 proposed by the other clients.
Also, the worker suggested by a client $P_j$ might collude
with $P_j$ against other clients (see Section~\ref{secpriv} on
collusions).
\end{note}  

\begin{algorithm}
\For(\tcc*[f]{EigenTrust-like global reputation}){$i=1$ \KwTo $n$}{
$P_i$ submits local reputation values $\{\ell_{ij}: j=1,\ldots,n\}$ 
to all $M$ accountability managers of $P_i$\;
\ForAll{pupils $P_d$ of $P_i$}{
$P_i$ collects local reputation values $\{\ell_{dj}: j=1,\ldots,n\}$\;
$P_i$ normalizes $c_{dj}=\ell_{dj}/\sum_j \ell_{dj}$, $j=1,\ldots,n$\label{norm};
}
\ForAll{pupils $P_d$ of $P_i$}{
\For{$j=1$ \KwTo $n$}{
 $P_i$ queries all the accountability managers of $P_j$ 
for $c_{jd}g_j^{(0)}$\label{acord};
}
$k:=-1$\;
\Repeat{$|g_d^{(k+1)}-g_d^{(k)}|<\epsilon$;\tcp*[f]{Parameter $\epsilon>0$ is a small value}} 
{
$k:=k+1$\;
$P_i$ computes $g_d^{(k+1)}=c_{1d}g_1^{(k)}+c_{2d}g_2^{(k)}+\ldots+c_{nd}g_n^{(k)}$\; 
\For{$j=1$ \KwTo $n$}{
$P_i$ sends $c_{dj} g_d^{(k+1)}$ to all $M$ accountability managers of 
$P_j$\;
$P_i$ waits for all accountability managers of $P_j$ to return $c_{jd}g_j^{(k+1)}$;
}
}
}
}
\caption{{\sc Co-utile P2P global reputation
computation}}
\label{globrep}
\end{algorithm}


\begin{algorithm}[]
Parameter $p \in [0,1]$\; 
\eIf{$P_s =  P_d$}{
$P_d$ decrypts $comp=SK_d(Ecomp)$\;
\eIf(\tcc*[f]{In this case, $msg$ contains an input}){$comp=nil$}{
$P_d$ decrypts $I||nonce=SK_d(msg)$\label{decrypt}\;
\lIf{$P_d$ has not previously received $nonce$}{$P_d$ appends $I$ to $Ilist$}
}
{
\eIf(\tcc*[f]{The worker has refused to compute}){$comp=\mbox{``refuse''}$}{$P_d$ assigns $out:=nil$;\tcc*[f]{The worker returns no output}}
{
$P_d$ waits until $Ilist$ contains $m$ different inputs\;
$P_d$ computes $out:=comp(Ilist)$\label{compute};\tcc*[f]{The worker computes}}
$P_d$ cleans $Ilist$\;
$P_d$ recovers $K=SK_d(msg)$\;
$P_d$ calls $\mbox{C-REV-CH}(P_d,E_K(out),Ecomp,P_{prev})$\label{endcompute};
}
}
{\leIf{$P_s$ is the originator of $msg$\label{2pas4}}{$p_{forward}=1$}{$p_{forward}=p$}
$P_s$ computes $decision=\mbox{Bernoulli}(p_{forward})$\label{2pas9}\;
\eIf(\tcc*[f]{$P_s$ decides to hop}){$\mbox{decision}=1$}{
$P_s$ sends $(msg,Ecomp)$ to $P_t=\mbox{\sc Select}(g_s,g_d)$\label{2pas16}
\tcc*[f]{See Note~\ref{select} about function {\sc Select}}\;
\eIf(\tcc*[f]{See Note~\ref{notflex} about $\delta$}){$P_s$'s reputation is at least $g_t - \delta$}{
$P_t$ calls $\mbox{C-FWD-CH}(P_t,msg,Ecomp,P_d)$\label{deal};
}
{
$P_t$ discards $(msg,Ecomp)$\label{discard}; 
}
}
{
$P_s$ directly sends $(msg,Ecomp)$ to $P_d$
\tcc*[f]{$P_s$ decides to submit to the worker}\;
$P_d$ decrypts $comp=SK_d(Ecomp)$\;
\If{$comp \neq nil$ and $P_s$'s reputation is less than $g_d - \delta$}
{
$P_d$ sets $Ecomp:=PK_d(\mbox{``refuse''})$;\tcc*[f]{$P_d$ refuses to compute}
}
$P_d$ calls $\mbox{C-FWD-CH}(P_d,msg,Ecomp,P_d)$;
}
}
\caption{$\mbox{C-FWD-CH}(P_s, msg, Ecomp, P_d)$}
\label{p2panrat}
\end{algorithm}

\begin{note}[On the flexibility parameter $\delta$]
\label{notflex}
In Protocol~\ref{p2panrat} (C-FWD-CH) a peer $P_j$ does not discard a message
from a peer $P_i$ as long as $P_i$'s reputation $g_i$ is at
least $g_j - \delta$, where $\delta$ is a small reputation amount.
The value $\delta$ introduces some flexibility in the interaction
and helps new peers (that start with 0 reputation) to earn
reputation as first forwardees. A large
$\delta$ is not acceptable from the rational point of view:
high-reputation peers have little to gain by accepting
messages or computations from peers much below them in reputation
(if those low-reputation peers are the clients, they may not
reward them).
\end{note}

\begin{note}[On function {\sc Select}]
\label{select}
In Protocol~\ref{p2panrat} (C-FWD-CH) function $P_t=\mbox{\sc Select}(g_s,g_d)$ is used by a peer $P_s$ 
to select a peer $P_t$ 
as a forwardee towards the worker $P_d$. There are several
ways in which this can be done. However, the rational choice
is for $P_s$ to select a forwardee $P_t$ with a sufficient reputation
so that $P_d$ does not refuse to compute should $P_t$ directly
submit to the worker. Hence, if $P_s$'s reputation is 
$g_s \geq g_d - \delta$, $P_s$ can randomly pick any of the 
peers whose reputation lies in $[g_d - \delta, g_s+\delta]$:  
any forwardee $P_t$ with reputation at least $g_d - \delta$ does not risk
refusal from the worker $P_d$, but peers $P_t$
with reputation above $g_s+\delta$ will discard $P_s$'s messages.
On the other hand, if $g_s < g_d - \delta$, $P_s$ chooses the peer
with the maximum reputation that does not exceed $g_s + \delta$, because
any peer with reputation above that value will discard $P_s$'s message. 
\end{note}


\begin{algorithm}[]
$P_s$ sends $(E_K(out),Ecomp)$ to $P_{prev}$\;
\eIf{$P_{prev}$ knows the key $K$}{
$P_{prev}$ decrypts $out$\label{3pas2}
\tcc*[f]{If $P_{prev}$ knows $K$, then $P_{prev}$ is the client and $P_s$ the first forwardee}\;
 $P_{prev}$ sends $S_{prev}(\mbox{``$P_{prev}$ has set $\ell_{prev,s}=1$''})$ to $P_s$ \label{commit}
\tcc*{$P_{prev}$ commits to rewarding the first forwardee}
$P_s$ forwards $S_{prev}(\mbox{``$P_{prev}$ has set $\ell_{prev,s}=1$''})$ to $P_{prev}$'s AMs\label{revfor}\;
$P_{prev}$'s AMs set $\ell_{prev,s}=1$\;
$P_s$ returns a signed reward receipt $S_s(\mbox{``$P_s$ acknowledges 
$\ell_{prev,s}=1$''})$ to $P_{prev}$\label{rewrec};
}
{
Let $P_{prev2}$ be the peer from whom $P_{prev}$
received the C-FWD-CH message with $Ecomp$\;
$P_{prev}$ calls $\mbox{C-REV-CH}(P_{prev},E_K(out), Ecomp, P_{prev2})$\label{endavant}.}
\caption{$\mbox{C-REV-CH}(P_s,E_K(out),Ecomp, P_{prev})$}
\label{p2prevrat}
\end{algorithm}

\begin{note}[On rewarding the first forwardee only]
\label{noterew} 
In Protocol~\ref{p2prevrat} (C-REV-CH) only
the first forwardee is rewarded, rather than all forwardees, and only
when the computation to be done is not $nil$. 
Note that in Step~\ref{norm} of Protocol~\ref{globrep} the local reputations
awarded by a peer are normalized before updating
global reputations (this is done to prevent peers from 
increasing their influence by giving more opinions on other peers).
Hence, if all forwardees were rewarded by the client, 
the reputation increase of each rewarded forwardee would be smaller.
Thus, every forwardee would be 
better off by sending $msg$ directly to the destination peer
 rather than forwarding it
to another forwardee.
As a consequence, there would be only one forwardee, who would know 
that the previous peer is the client who originated $msg$. 
This would break the anonymity of the channel.
Rewarding only the first forwardee when the computation to be done
is not $nil$ avoids this problem 
and is a sufficient incentive: the forwardees do not see
which is the computation to be performed (it is probabilistically encrypted
under the worker's public key) and any
 forwardee can hope to be the first for a non-nil computation 
and thus has a reason
to collaborate.
\end{note}

\section{Co-utility analysis}
\label{discoutil}

We argue that the framework formed
by Protocols~\ref{pro1rat},~\ref{globrep},~\ref{p2panrat} and~\ref{p2prevrat} is co-utile,
that is, that these protocols will be adhered to by the rational players.
We first give a sketch justification and then we analyze in detail  
the motivation of each of the above 
player categories to adhere to the protocols they are required to
participate in. The sketch is as follows:
\begin{itemize}
\item The clients' goal is to perform a joint computation and obtain
their respective correct outputs, while keeping their own inputs and 
outputs private. For that reason, the clients can be assumed to 
correctly perform their tasks in Protocols~\ref{pro1rat},~\ref{p2panrat} 
and~\ref{p2prevrat}.
If a client fails to behave as prescribed (when acting as a client,
as a forwardee or as a worker), her reputation will decrease
and it will be 
more difficult for him/her to obtain correct results (see 
Section~\ref{discoutil}).
\item Forwardees have no role in Protocol~\ref{pro1rat}, 
but they are essential to the operation of the co-utile anonymous channel in 
Protocol~\ref{p2panrat} (C-FWD-CH) and 
Protocol~\ref{p2prevrat} (C-REV-CH). Their incentive is the hope to be rewarded
in Protocol~\ref{p2prevrat} with a reputation increase in 
case they turn out to be the first forwardee.
\item Workers are expected in C-FWD-CH to perform the required computation. Then 
in C-REV-CH they are expected to start the reverse path upstream to return the output.
Their incentive to compute correctly is to be rewarded 
in Protocol~\ref{pro1rat} if the output they 
deliver is the majority output among those returned by redundant workers.
\item Accountability managers have important roles in 
Protocols~\ref{pro1rat},~\ref{globrep} and~\ref{p2prevrat}.
In our security model (Section~\ref{players}),
 peers are assumed to be rational,  
even if rational attackers are not excluded.
Given that peers interact in successive iterations, the interest
of rational accountability managers is to favor
correct computations, as they may be clients themselves in subsequent
computation rounds.
On the other hand, the fact that the $M$ accountability managers of every peer
are (pseudo)randomly assigned thwarts conflicts of interest
and facilitates honest management of the peer's reputation.
Furthermore, if necessary, a countermeasure could be added 
right after Step~\ref{acord} of Protocol~\ref{globrep} 
whereby $P_i$ punishes with local reputation 0 those AMs that
provide local reputations for $P_j$ that do not agree with 
the majority reputation value.
\end{itemize}

Next, we elaborate on co-utility for each player category.

\subsection{Co-utility for clients}

In Protocol~\ref{pro1rat}, $ID_1,\ldots, ID_m$ can be 
assumed to honestly 
agree on the joint computation, because jointly 
computing is their goal.  Then at Step~\ref{pseudo} 
they are also interested in switching
to their respective pseudonyms $P_1, \ldots, P_m$.
If a client used her real identity $ID_i$, in C-FWD-CH it
would be trivial for a forwardee $P_t$ to know whether she is the 
first forwardee (and hence the only forwardee that will be rewarded);
in consequence, if $P_t$ was asked to be a forwardee
by a pseudonymous peer, $P_t$'s rational decision would be to decline.
In this way, there would be only one forwardee, which would weaken
unlinkability for the clients.

Also in Protocol~\ref{pro1rat} it is rational for the 
pseudonymous clients to correctly perform steps up to
Step~\ref{paspas}. This means pruning $C$ into their
respective computations,
 selecting workers, and sending to the workers via
the anonymous channel C-FWD-CH first their private
inputs and then the pruned computation and the 
key for receiving encrypted outputs. 
Note that private inputs are sent {\em in parallel} 
at Step~\ref{inputsend}, so $P_i$ cannot wait for the 
other peers to send their private inputs to $P_i$'s workers and
then free-ride without sending $P_i$'s input to the other
peers' workers.
It is in all the clients' interest to perform honestly, as they want to obtain
correct outputs.

In Protocol~\ref{p2panrat}, it is bad for the client 
$P_i$ originating the message (called $P_s$ inside the protocol) 
to directly send it to the worker $P_{i_k}$ (called $P_d$ inside the protocol), 
even if only
with probability $1-p$, like in the Crowds system~\cite{crowds}.
In that case, from the worker $P_{i_k}$'s viewpoint, the most likely sender
 of a message is the originating client $P_i$: indeed, $P_i$ 
would send it with probability $1-p$, 
 whereas 
the $l$-th forwardee would send it 
only with probability $p^l (1-p) < 1- p$. Hence, the originating
client $P_i$
is interested in looking for a forwardee, and therefore $P_i$ will honestly
follow Steps~\ref{2pas4} and~\ref{2pas9}. Further, the client $P_i$
wants, if possible, a forwardee that will not risk refusal by the worker
or, if this is not possible, a forwardee with the maximum
possible reputation among those that will not discard 
$P_i$'s message (see Note~\ref{select}
for a justification).
{\em Since a client $P_i$ will only obtain her output $O_i$ 
if she finds her own good forwardees and workers (the forwardees
and workers of other clients do not take care of $O_i$), 
$P_i$ is motivated to maintain a high reputation $g_i$.}
An additional motivation for a client $P_i$ to maintain a high reputation
is that it allows $P_i$ to randomly 
choose her first forwardee among a large set of peers that will not risk
refusal by the worker, which increases unlinkability of successive
messages sent by $P_i$.
This ensures honest adherence to 
Step~\ref{2pas16}.

In Protocol~\ref{p2prevrat},
it is obviously in the interest of $P_i$ (called $P_{prev}$ inside
the protocol) to decrypt her private output when she
receives it (Step~\ref{3pas2}). 
Further $P_i$'s best option is to reward the first forwardee by giving
her local reputation 1 (Step~\ref{commit}), in order to obtain a reward
receipt from the first forwardee (Step~\ref{rewrec}). This reward receipt
is necessary for $P_i$ to escape being punished with 0 local reputation in 
Step~\ref{1pas14} 
of Protocol~\ref{pro1rat}.

$P_i$ could certainly decide to favor a false first forwardee $P'$
of her choice, rather than the real first forwardee $P$.
This would still work well for $P_i$, because
$P'$  would return a signed receipt for the same reasons
that $P$ would do it.
However, if $P_i$ wants to favor $P'$, it entails
less risk (of being discovered)
for $P_i$ to use $P'$ as a {\em real} first forwardee.
Thus, there is no rational incentive for clients to favor
false first forwardees.


\subsection{Co-utility for forwardees}

In Protocol~\ref{p2panrat}, there may be two types of forwardees:
\begin{enumerate}
\item $P_s$ is a forwardee if she is not the originator of $msg$.
$P_s$'s incentive to perform her role properly is the hope of being
the first forwardee {\em for a non-nil computation} 
and hence be rewarded with a reputation increase. Note that
$P_s$ does not know whether she is the first and whether
the computation is non-nil, because the latter is encrypted and 
timing does not help, given that the C-FWD-CH calls with nil and
non-nil computations are interleaved due to the hopping mechanism
among forwardees.
According to the protocol, $P_s$ must decide between forwarding 
$(msg,comp)$ to some other peer $P_t$ or directly sending 
$(msg,comp)$ to the destination peer $P_d$ who will perform $comp$.
Both actions take about the same effort, so it is rational 
 for $P_s$ to make the decision randomly according
to the prescribed probabilities ($p$ for forwarding and $1-p$ for
directly sending to $P_d$). 
In case of forwarding, $P_s$'s rational procedure
is like the originator's: use the {\sc Select} function.
\item $P_t$ is the forwardee selected by $P_s$ in case of forwarding.
$P_t$ must decide
on message acceptance or discarding. The incentive for $P_t$ to accept
to deal with a message is the hope to increase her reputation if 
$P_t$ happens to be the first forwardee 
for a non-nil computation (which $P_t$ does not know).
 However, $P_t$ will not accept to deal with a message coming
from $P_s$ if the latter's reputation is too low: it might be a sign
that $P_s$ did not ``pay'' previous first forwardees in Step~\ref{commit}
of Protocol~\ref{p2prevrat} and was therefore punished with reputation
decrease in Step~\ref{1pas14} of Protocol~\ref{pro1rat}. Thus,
the rational decision is for $P_t$ to accept to deal with $P_s$'s message
only if $P_s$ is not too inferior to $P_t$ in 
reputation, that is, if $g_s \geq g_t - \delta$.
Otherwise, in Step~\ref{discard} 
$P_t$ discards $P_s$'s message.
\end{enumerate}

In Protocol~\ref{p2prevrat}, forwardees backtrack  along the reverse
path from the worker to the originator. There are two forwardee roles:
\begin{enumerate}
\item In Step~\ref{revfor} $P_s$ is the first forwardee
because $P_{prev}$ was the originator under C-FWD-CH. Hence, at Step~\ref{revfor}
$P_s$ is obviously
interested in forwarding to $P_{prev}$'s AMs the reward $P_s$ receives from $P_{prev}$.
On the other hand, at Step~\ref{rewrec}
$P_s$'s best option is to return the reward receipt to $P_{prev}$, 
because $P_{prev}$ could otherwise blacklist $P_s$
and never make $P_s$ a first forwardee in future joint computations.
\item In Step~\ref{endavant} $P_{prev}$ does not know whether she
is the first forwardee or not.
In the hope of being the first forwardee, $P_{prev}$'s best option
is to forward $(E_K(out),comp)$ to the next upstream peer $P_{prev2}$. 
If $P_{prev2}$ turns out to be the originator, then $P_{prev}$ will be rewarded
as the first forwardee (see previous item).
\end{enumerate}

\subsection{Co-utility for workers}

In Protocol~\ref{p2panrat} at Step~\ref{decrypt} $P_d$ is the 
worker to whom the private inputs are sent by the clients. 
Decrypting and storing these inputs is necessary
for $P_d$ to be able to correctly perform the computation at Step~\ref{compute}.
The motivation for $P_d$ to return the correct computation result
encrypted under the retrieved key $K$ 
in Step~\ref{endcompute} is to receive a reputation reward in 
Step~\ref{reprew} of Protocol~\ref{pro1rat}, rather
than a reputation punishment in Step~\ref{reppun} of that protocol. 

In Protocol~\ref{p2prevrat}, the worker $P_d$ (called $P_s$ within 
the protocol) merely sends $(E_K(out),Ecomp)$ to the previous peer
in the path followed by C-FWD-CH. The worker's motivation to do this 
is to earn a reputation reward in
Step~\ref{reprew} of Protocol~\ref{pro1rat}.

Note that, workers being peers, they are also likely to be clients
at some point. And for a client having a high reputation is the way
to easily find forwardees that help preserve the client's 
input and output privacy.

\subsection{Co-utility for accountability managers}

In Protocol~\ref{pro1rat}, the accountability managers
punish those clients that do not reward their first 
forwardee (Step~\ref{1pas14}). 
Since they are pseudorandomly chosen, most AMs are 
rational and thus interested in favoring correct computation; hence,
most AMs will discharge their role as described in the protocol.

Protocol~\ref{globrep} is run by all peers in their capacity of accountability
managers. Since most AMs are rationally
interested in favoring correct computation, they are also interested
in correctly updating 
and maintaining global reputations. Therefore, most AMs can be
assumed to run Protocol~\ref{globrep} correctly. 
For the same reason, all peers will supply local reputations
to their AMs at the start of the protocol.

Finally, in Protocol~\ref{p2prevrat} the role of AMs is to 
reflect the reputation rewards received by the first forwardees.
Again, the rational interest of the AMs to preserve the correct
operation of the protocol suite allows expecting them to do their job.

The bottom line that allows trusting AMs as a community is that they are
pseudorandomly chosen and redundant (each peer has $M$ AMs), coupled
with the assumption that a majority of peers (and hence of AMs) is rational. 
Yet, it might be argued that assuming a 
majority of rational peers is not the same
as assuming a majority of honest peers. If it is feared
that a sizeable proportion of AMs might have rational motivations
to deviate,  additional countermeasures could be set up 
to punish bad AM behavior.
For example, as hinted earlier in the sketch at the beginning
of this section,
after Step~\ref{acord} of Protocol~\ref{globrep}
$P_i$ could punish with local reputation 0 those AMs that
provide local reputations for $P_j$ that do not agree with
the majority reputation value. Yet, to avoid complicating 
pseudocodes,
we have refrained from including such optional countermeasures.

\section{Privacy and correctness}
\label{privsec}

Here we examine how the protocols satisfy the 
requirements of Section~\ref{requirements}, and thereby
preserve the confidentiality of the peers' private inputs and outputs
and incentivize correct computation by the peer workers.

\subsection{Privacy}
\label{secpriv}

In our protocol suite, the privacy of client inputs and outputs is based
on the co-utile anonymous channel implemented by protocols
C-FWD-CH (downstream) and C-REV-CH (upstream). This channel breaks
the link between clients and their inputs and outputs.

We first show that, in general, no collusion of peers can link
with certainty an input to the corresponding client. 
Then we deal with the unlinkability of outputs and 
we argue that the only collusion that could link
an output to the corresponding client is hardly rational.
After showing that collusions are improductive to link inputs
and either improductive or not rational to link outputs, 
we prove that in the absence of collusion 
a client can plausibly deny that a certain input is hers
and that a certain output is hers.

\begin{prop}
\label{propcoll}
If a client $P_i$'s input $I_i$ is not embedded 
in her computation $C_i$, no collusion of peers can link with certainty an
input to $P_i$.
\end{prop}

\begin{IEEEproof} A successful collusion must include at least
a first forwardee (who knows the client's pseudonym)
and a worker (who
computes the client's output).
Under the assumption of the proposition, client $P_i$
 sends her input only at 
Step~\ref{inputsend} of Protocol~\ref{pro1rat}, via C-FWD-CH. In this call
to C-FWD-CH,  $P_i$
does not reward the first
forwardee, which means that the latter does not learn
 she is the first forwardee. Hence the worker cannot 
find a first forwardee that knows for sure the pseudonym
of the client to whom a specific input corresponds.
\end{IEEEproof} 

Regarding outputs, when a client $P_i$ calls C-FWD-CH to send the return
key and the computation (Step~\ref{paspas} of Protocol~\ref{pro1rat}), 
$P_i$ does reward the first forwardee.
Therefore, this first forwardee learns $P_i$'s pseudonym
and could collude with the worker to link the output
to the client's pseudonym.
Nonetheless, such a collusion is hardly rational:
\begin{itemize}
\item Peers are pseudonymous: the first forwardee
only knows the worker's pseudonym and the worker
does not even know the first forwardee's pseudonym
unless the first forwardee tells the worker.
People tend to collude with those they know, 
and thus pseudonymity is a defense against collusion.
\item There is an asymmetry between the worker and the first forwardee. 
It is unclear why 
the worker should share with the colluding first forwardee
the value of the client's private output after 
the first forwardee reveals the client's pseudonym:
the worker cannot verify that the pseudonym actually corresponds to the
client, and even if the worker accepts the pseudonym as good, 
it is not the client's real identity. 
\item In case a first forwardee or a worker $P_t$ is also a client herself,
colluding to link another client $P_v$'s output with $P_v$'s pseudonym
makes $P_t$'s own output more easily linkable (linkage possibilities
reduce from $m$ clients to $m-1$ after a successful collusion).
\end{itemize}
A final mitigating factor is that, 
in many joint computations the outputs are either not private 
({\em e.g.} the tally in e-voting is public) 
or less confidential than the inputs.

\begin{prop} 
\label{proppriv}
If there is no collusion between the  
peers, a client $P_i$ can plausibly deny 
that a certain private input is hers and 
can also plausibly deny that a certain private output is hers.
Hence, client privacy would still 
hold if the mapping between the client's pseudonym 
$P_i$ and the client's
real identity $ID_i$ was discovered. 
\end{prop}

\begin{IEEEproof}The privacy guarantee is based on unlinkability
and input/output encryption. 

Let us examine whether a worker
can link an input to the corresponding client. 
By the design of Protocol C-FWD-CH, a worker $P_{j_k}$ knows that
when $P_{j_k}$ receives $PK_{j_k}(I_i)$ from a peer $P$, $P$
is unlikely to be the client $P_i$ to whom $I_i$ corresponds (a client 
always chooses to hop if she can). Yet, since we assume in 
Protocol~\ref{pro1rat}
that $m \geq 4$, there are at least 4 clients;
hence, even if both $P_{j_k}$ and $P$ happened to be 
also clients themselves,  $P_{j_k}$
cannot unequivocally determine which of the remaining clients is $P_i$.
What $P_{j_k}$ can do is to estimate the probability 
that $PK_{j_k}(I_i)$ was submitted by the $l$-th forwardee as $p^{l-1}(1-p)$,
and hence that the most likely submitter is the first forwardee.
Nevertheless, the first forwardee is chosen by client $P_i$ 
using the {\sc Select} function, described in Note~\ref{select}.
If $g_i \geq g_d - \delta$, then $P_i$ chooses the first forwardee
randomly among the set of peers with reputation in 
the interval $[g_d-\delta, g_i + \delta]$,
and this set depends on the current reputations and varies over time;
hence, as long as there are several peers with reputations in 
the previous interval,
even if $P_{j_k}$ received two encrypted inputs from the same peer
in two different successive joint computations, $P_{j_k}$ cannot
infer that both inputs correspond to the same client. 
If $g_i < g_d - \delta$, then $P_i$ chooses as a first forwardee
the peer with the maximum reputation that does not exceed 
$g_i + \delta$: if reputations do not change between two 
successive messages, $P_i$ would choose the same first forwardee
for both messages; yet $P_d$ cannot be sure that the submitter
of both messages is really the first forwardee, and hence 
$P_d$ cannot be sure that both messages were generated
by the same client. Hence, in no case can two different messages
sent by the same client be unequivocally linked, even if the 
probability of correctly linking them is lower when $g_i \geq g_d - \delta$.

On the other hand, the worker $P_{i_k}$ 
receives the computation $C_i$ via C-FWD-CH from the same client $P_i$ to whom 
output $O_i$ must be returned via C-REV-CH. 
However, by the design of C-FWD-CH $P_{i_k}$ receives $C_i$ 
without learning who $P_i$ is. Also, by the design of C-REV-CH
$P_{i_k}$ returns the client's output $O_i$ by hopping upstream via
the reverse path, with no knowledge of $P_i$'s pseudonym or identity.
Certainly,  $P_{i_k}$ could try to identify the client $P_i$ 
by looking for which peers $P_{i_k}$ is closest in reputation; 
however, $P_{i_k}$ does not know the parameter $\kappa_i$ (number
of closest peers) used by $P_i$ in Protocol~\ref{pro1rat} when choosing 
her workers.

Consider now linkability by a forwardee $P_s$. 
If $P_s$ is a forwardee of 
a message from $P_{prev}$, in general $P_s$ does not
know whether $P_{prev}$ originated the message
 or is merely
forwarding it. The only exception is when $P_s$
is the first forwardee 
 (because in this case $P_s$ receives a message from $P_{prev}$
 in Step~\ref{commit} of Protocol~\ref{p2prevrat}).
Yet, in this case $P_s$ can only link the {\em encrypted} 
version of the output (that is, $E_K(out)$) to the client's identity 
$P_{prev}$.

Lastly, if $AM$ is an accountability manager of a client $P_i$,
$AM$ only sees the reward receipts from the first forwardees
of that client (see Protocol~\ref{pro1rat}).
In no case can $AM$ access the inputs
or the outputs of $P_i$. 

Since no input or output can be unequivocally 
attributed to a client by any other peer, plausible
deniability by the client holds.
\end{IEEEproof} 

\subsection{Correctness}

To ensure correctness, the delivery of correct outputs
must be 
the best rational option for the non-client peers (forwardees and workers). 
Protocols~\ref{pro1rat}, \ref{p2panrat} and~\ref{p2prevrat} are designed to
incentivize this option. Thus, from the discussion on co-utility 
for workers and forwardees in Section~\ref{discoutil} 
the following proposition follows:

\begin{prop}
\label{propsecu}
The rational behavior of workers and forwardees in 
Protocols~\ref{pro1rat}, \ref{p2panrat} and~\ref{p2prevrat} is to return
correct outputs to the clients.
\end{prop}

\section{Example applications}
\label{applications}

Our framework is totally general and therefore it can
accommodate any joint computation with input and output confidentiality.
However, for the sake of illustration,
we sketch 
two specific applications.

\subsection{Finding differences with 
the previous and following parties in a ranking or an auction}

In this application, each client $P_i$ gives as private input $I_i$ 
her value for a certain magnitude ({\em e.g.} wealth or auction bid) and 
wants as private output $O_i$ 
 the difference with the previous party in the ranking
by that magnitude ({\em e.g.} how richer is
the next richer client or how much more did the next higher bidder offer) 
and the difference with 
the following party in the ranking
({\em e.g.} how poorer is the next poorer client or how much less
did the next lower bidder offer).

In this case, the computation $C$ consists of ranking 
all clients and finding the differences between successive clients
in the ranking. The pruned computation $C_i$ of interest for $P_i$ 
consists of ranking all clients and finding only 
the differences between $P_i$ and its previous client
and between $P_i$ and its following client. Hence, in this application,
$P_i$'s input $I_i$ must be embedded in $C_i$ (yet this is no problem,
because the worker running $C_i$ cannot link $I_i$ to $P_i$). 

%
%
%
\subsection{Electronic voting}

In electronic voting, the clients $P_1, \ldots, P_m$ are the voters.
The private inputs of the clients $I_1, \ldots, I_m$ are their secret
votes. There is only one output $O$, the tally, which is public. 
Note the value of a vote does not reveal the identity of the 
voter, and hence it is safe for clients to send their 
votes to workers as long as they do it using an anonymous channel.

The computation $C$ consists of computing the absolute 
frequencies of all options 
susceptible of being chosen by voters ({\em e.g.} candidates,
parties, Yes/No/Blank etc.).
In this case $C_1=C_2= \ldots =C_m=C$ and 
no client input needs to be embedded in $C$. Rather
than each worker returning $O$ to her client via C-REV-CH,
it is simpler for all workers
to publish $O$.
If a worker publishes a tally $O'$ that disagrees with the majority
tally, $O'$ will be discarded (and the worker's reputation will decrease).
In fact, the computational overhead can be strongly reduced 
with the following simplification: 
each client chooses only {\em one} worker rather than $r$; 
in this case, there is a set of only $m$ workers (rather than
$m \times r$), but if $m \geq 4$, 
this set is enough to compute $O$ as the majority 
output.

It is also interesting to consider the particular case in which 
all peers are voters/clients. 
Since all peers are clients and 
they are rational, correct computation of the common
output (tally) interests them. 
Hence, most peers can be expected to behave honestly in their
capacity of workers. As a consequence,
{\em reputations are not needed} for the correct tally 
 to be majority.

We can generalize the last remark into the following proposition.

\begin{prop}
\label{propcommon}
If all peers are rational clients and there is a single common output
of the joint computation, Protocols~\ref{pro1rat},~\ref{p2panrat} 
and~\ref{p2prevrat} can be expected to yield a correct output even if
all operations related to reputations are suppressed from the protocols. 
\end{prop}


\section{Experimental results}
\label{experiments}



In this section, we report the results of the experiments with
the proposed co-utile framework. We compared with a {\em baseline} framework
in which there is worker redundancy but {\em no reputation} 
(that is, Protocols~\ref{pro1rat}, \ref{p2panrat} and~\ref{p2prevrat} 
without using reputations to decide on workers, forwardees or clients).
We considered good peers (who honestly do their job) and malicious peers
(who do not).
 
{\em Expected behavior.} If the co-utile framework is well designed, 
good clients (that are good peers) 
should obtain a higher rate of correct outputs in
it than in the baseline framework. 
Also, in the co-utile framework the rate of correct outputs for 
good clients should be higher than for bad clients (who are malicious peers).
Further, the reputation of the peers should highly and positively correlate 
with their goodness and with the rate of correct outputs they obtain. 
The rationale is that peers who perform wrong computations as workers 
are likely to end up having low reputation, and it is difficult for a 
peer with low reputation to find (as a client) other peers 
that are willing to perform a computation for her (as workers) 
or forward the former peer's messages (as forwardees).

{\em Experimental setting.} We built a P2P network with $n=100$ peers 
and we let it evolve for 250 iterations; in each iteration,
a joint computation was conducted by $m=10$ clients randomly chosen
among the 100 peers. Each client gave a numerical value as a private 
input to the joint computation; the latter consisted of ranking the input values
and returning as a private output the rank of the client's value.
We set $r=3$ (three redundant workers per client) and we ran 
simulations for different proportions of malicious peers. A malicious 
worker always returned a random output instead of the true output
of a computation. In the baseline framework {\em no} reputation 
management was used by peers to choose workers or forwardees,
or to decide whether to act as a worker or a forwardee. 
In the co-utile framework, we gave an initial reputation $1/n=0.01$
to each client and we took $\delta=0.002$. 

\begin{figure}
\begin{center}
\includegraphics[width=0.45\textwidth]{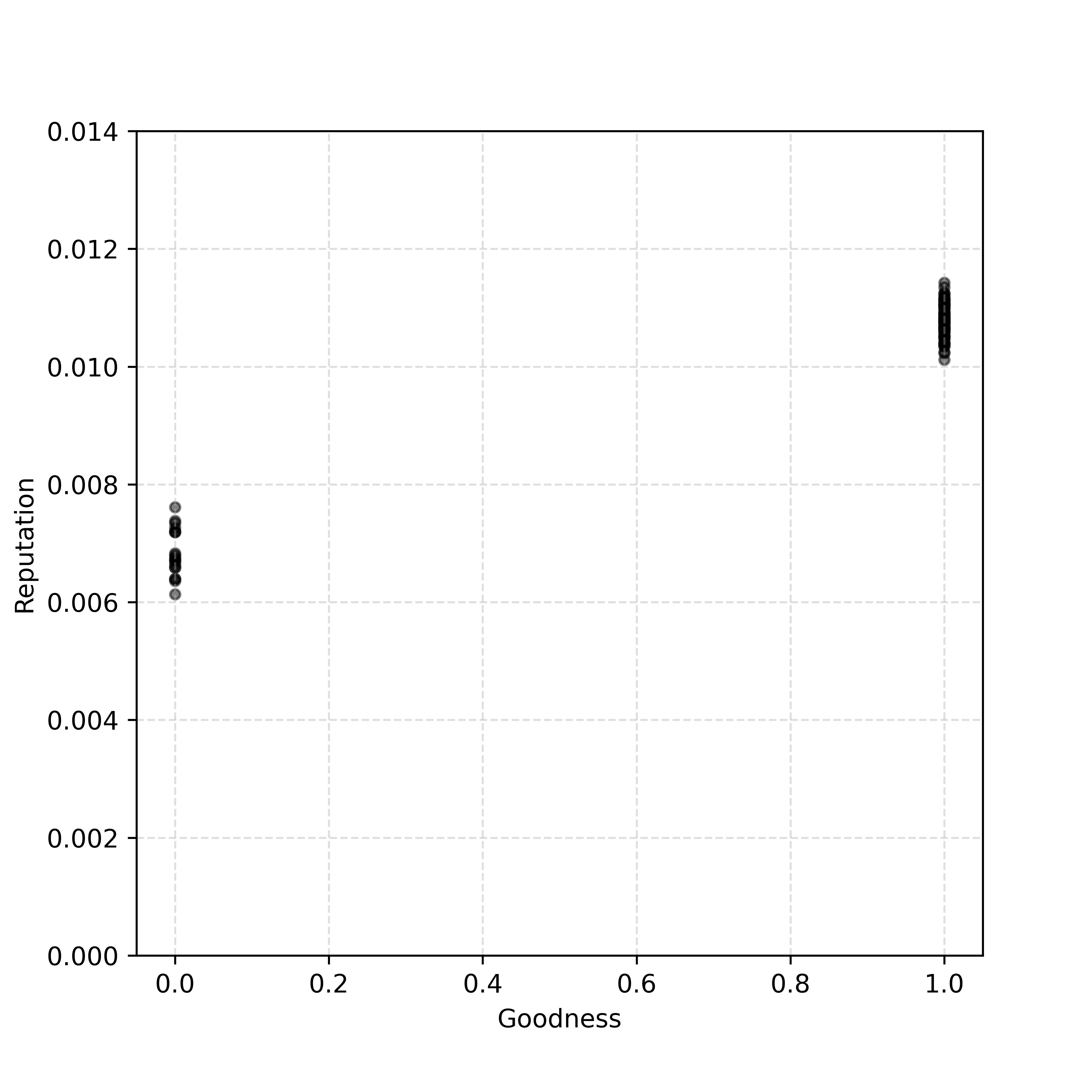}
\end{center}
\caption{Reputation of peers as a function of their goodness,
after 250 iterations, with $\delta=0.002$ and 20\% of malicious peers}
\label{goodrep}
\end{figure}

Figure~\ref{goodrep} plots the goodness of 
each peer (probability of behaving honestly)
against her attained reputation after the 250 iterations, for 
a 20\% proportion of malicious peers. It can be seen
that both are very highly correlated: all malicious peers
(those with probability zero of honest behavior) ended up
with reputations between 0.006 and 0.008, whereas 
all good peers (those with probability 1 of honest behavior)
attained higher reputations, between 0.010 and 0.012. Thus 
reputation captured very well the behavior of peers.

\begin{figure}
\begin{center}
\includegraphics[width=0.45\textwidth]{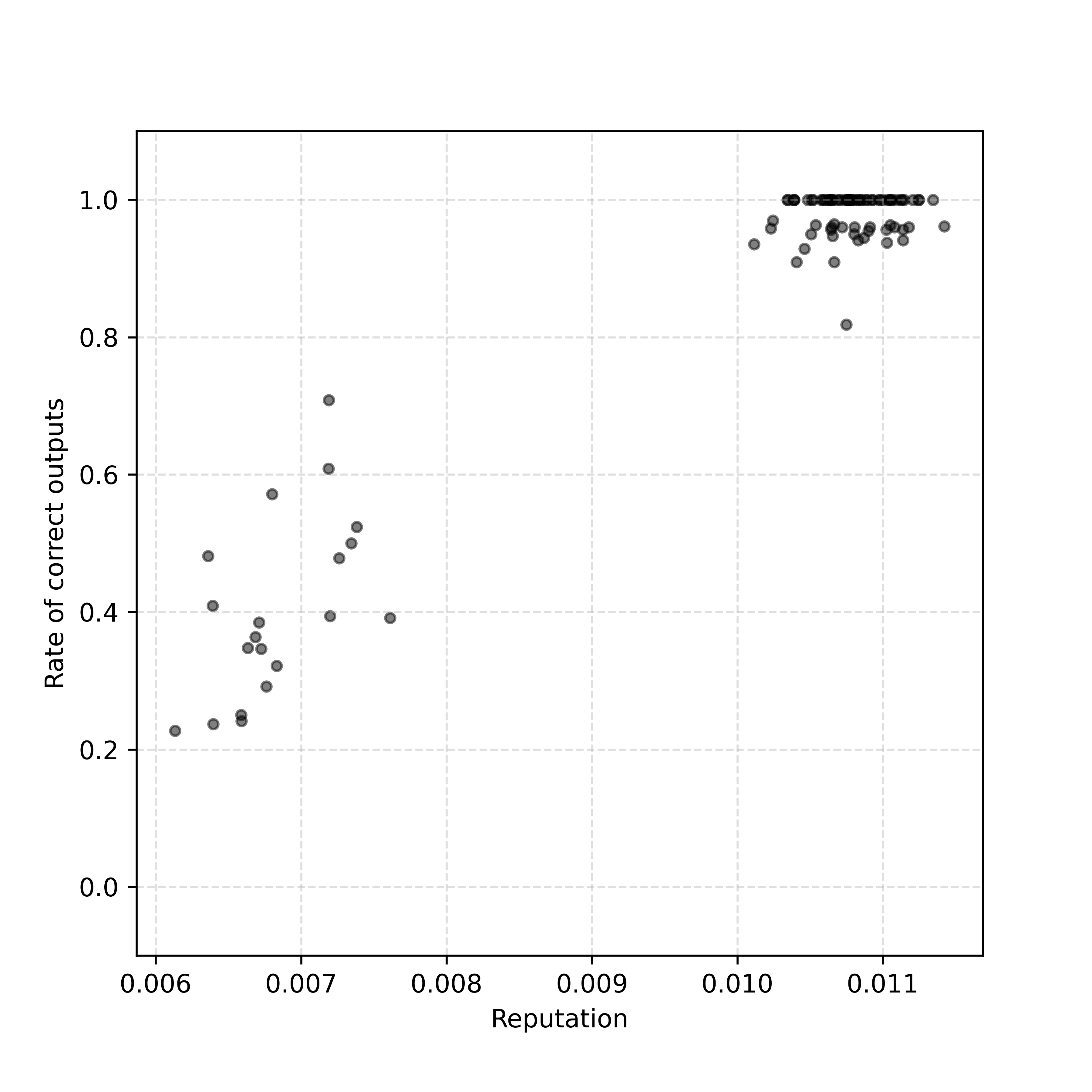}
\end{center}
\caption{Rate of correct outputs as a function of the reputation
of clients, after 250 iterations, with $\delta=0.002$ and 20\% of malicious peers}
\label{rate250}
\end{figure}

\begin{figure}
\begin{center}
\includegraphics[width=0.45\textwidth]{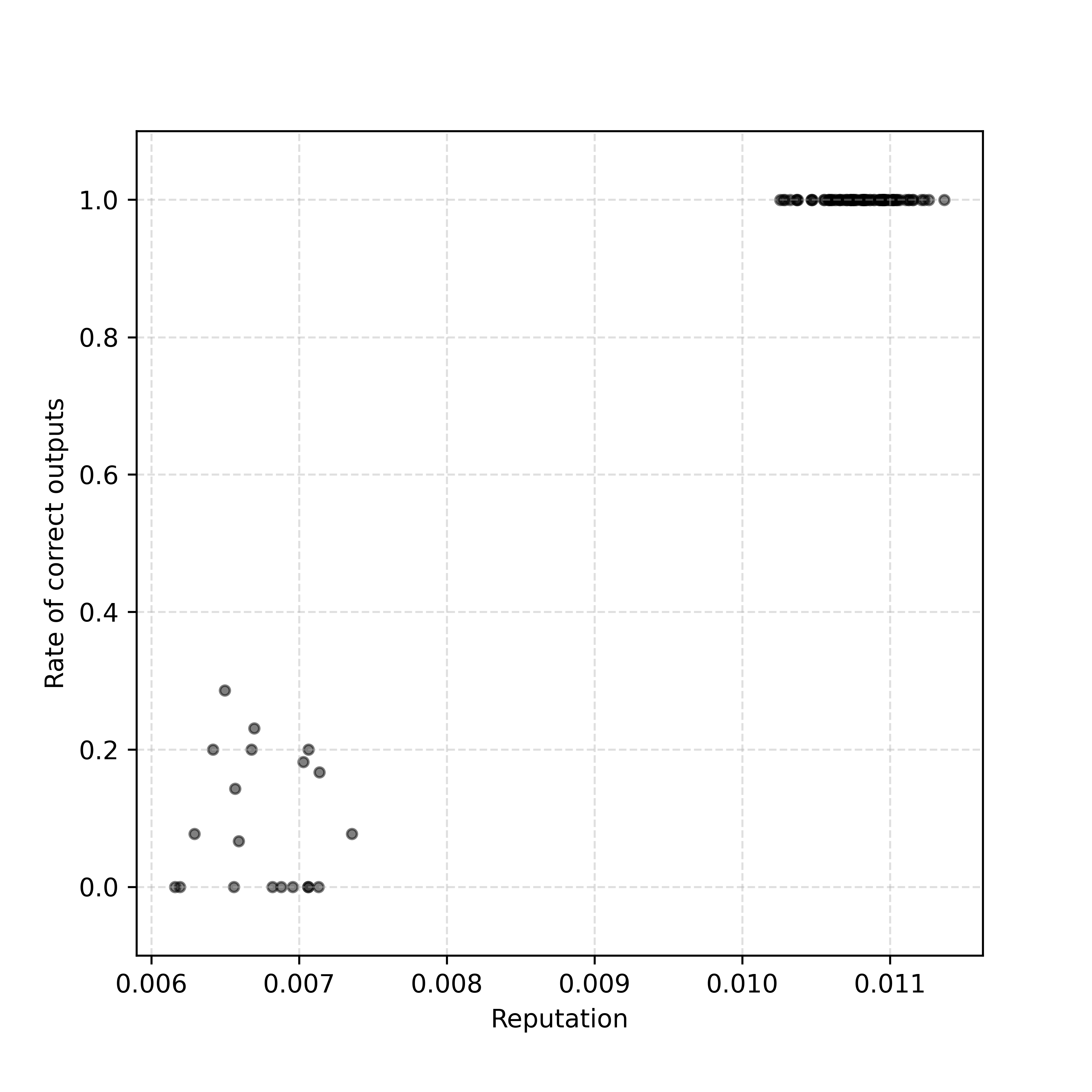}
\end{center}
\caption{Rate of correct outputs as a function of the reputation
of clients in the last 100 iterations, 
with $\delta=0.002$ and 20\% of malicious peers}
\label{rate100}
\end{figure}

Figure~\ref{rate250} displays the rate of correct outputs as a function
of the reputation of clients, both magnitudes accumulated 
after 250 iterations, for a 20\% proportion of malicious 
peers and $\delta=0.002$. Note that after 250 iterations all peers 
had been clients.
Nearly all the outputs requested by clients with high reputation were 
correct, whereas on average less than 50\% of the outputs requested by 
clients with low reputation were correct. This satisfied 
our expectations on the behavior of the co-utile framework.
In fact, as it can be seen in Figure~\ref{rate100}, reputation was 
even more decisive
in the last 100 iterations. In these last iterations, 
all computations requested by
peers with high reputation were correct, whereas on average 
less than 20\% of the 
outputs requested by clients with low reputation were correct. Hence, as soon the
system stabilizes, the good peers {\em always} obtain correct outputs.

\begin{figure*}
\begin{center}
\includegraphics[width=\textwidth]{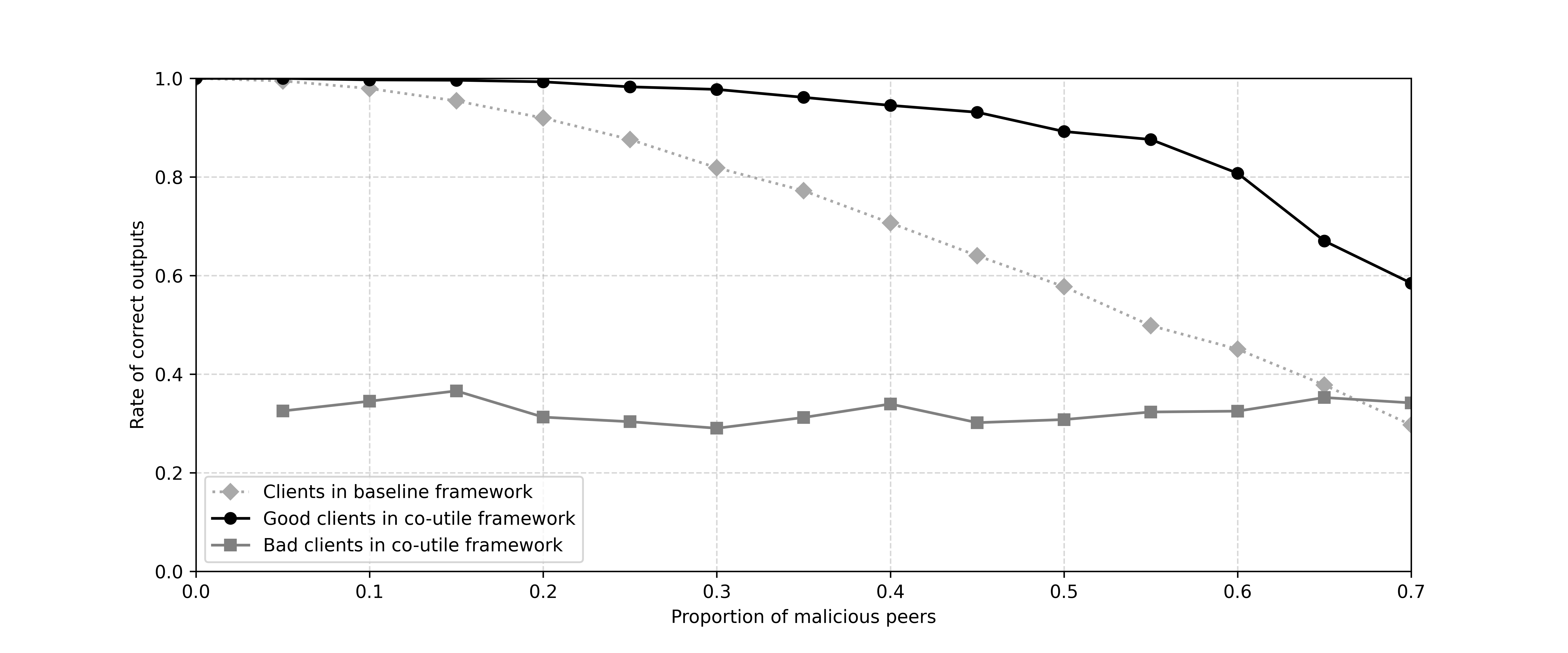}
\end{center}
\caption{Rate of correct outputs in the baseline 
and the co-utile frameworks for different types of clients, 
as a function of the proportion of malicious peers}
\label{corrmal}
\end{figure*}

Finally, Figure~\ref{corrmal} shows the rate of correct outputs in 
the baseline and the co-utile frameworks as a function of the proportion 
of malicious peers, after 250 iterations.
 No matter the framework, the rate of correct
outputs decreases as the proportion of malicious peers increases, which
was to be expected. However, in the co-utile framework good clients obtained a 
much higher rate of correct outputs than bad clients. Furthermore, good clients
in the co-utile framework obtained a higher rate of correct outputs 
than clients in the baseline framework, whereas bad clients in the co-utile
framework obtained a {\em much} lower rate of correct outputs
than clients in the baseline framework. 
Hence, reputation management in the co-utile framework is useful
to discriminate between good and bad clients, which in fact encourages
rational behavior: honest behavior turns out to be the best
rational option for any peer that wants to become a client. 

\section{Conclusions and future work}
\label{conclusions}

We have presented a peer-to-peer framework for multiparty computation
that has the following innovative features: (i) it is general-purpose
without making use of circuits, so that it can work for any computation
expressed as ordinary high-level programming code, no matter its
complexity, loops or recursions; 
 (ii) unequivocal linkability  
of each party's inputs is prevented; (iii) if a majority of 
peers are rational (not necessarily semi-honest), collusion 
is unattractive and hence, unequivocal linkability of outputs is also 
prevented; 
(iv) the rational behavior of peers is to return correct outputs
to the client parties.

Future research will be devoted to reducing the overhead 
caused by worker and accountability manager redundancy while 
preserving the current privacy and correctness guarantees. 
Further work will be performed on parameter tuning, and in particular 
on the value of the initial reputation to be assigned to newcomers.

\section*{Acknowledgments and disclaimer}

Thanks go to Oriol Farr\`as for comments on a 
draft of this paper.
We acknowledge support from the European Commission
(projects H2020-871042 ``SoBigData++'' and H2020-101006879 ``MobiDataLab''),
the Government of Catalonia (ICREA Acad\`emia Prize to the first author 
and grant 2017 SGR 705),
and the Spanish Government (project RTI2018-095094-B-C21
``Consent'').
We are with the UNESCO Chair in 
Data Privacy, but the views in this paper are our own and not
necessarily shared by UNESCO.

\begin{IEEEbiography}[{\includegraphics[width=1in,height=1.25in,clip,keepaspectratio]{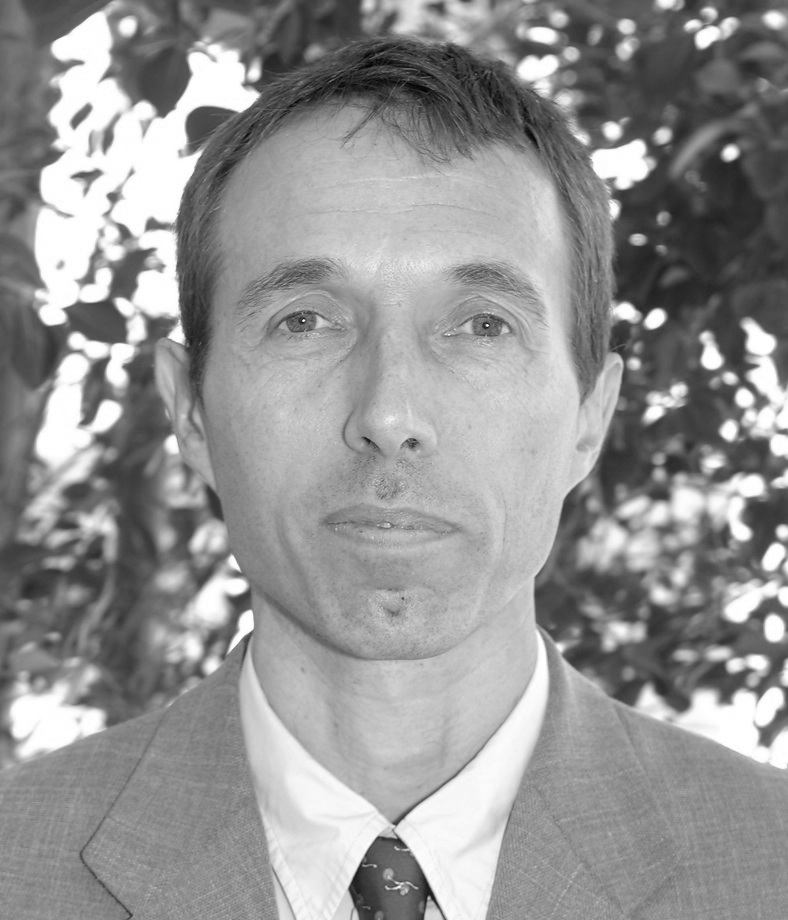}}]{Josep Domingo-Ferrer}
(Fellow, IEEE)
is a Distinguished Professor of Computer Science and
an ICREA-Acad\`emia Researcher at Universitat Rovira i Virgili,
Tarragona, Catalonia, where he holds the UNESCO Chair in Data Privacy
and leads CYBERCAT (Center for Cybersecurity Research of Catalonia). 
He received the MSc and
PhD degrees in Computer Science from
the Autonomous University of Barcelona in 1988 and
1991, respectively. He also holds an MSc degree in
Mathematics.
His research interests are in data privacy, data security and cryptographic
protocols. More information on him can be found
at \url{http://crises-deim.urv.cat/jdomingo}
\end{IEEEbiography}

\begin{IEEEbiography}[{\includegraphics[width=1in,height=1.25in,clip,keepaspectratio]{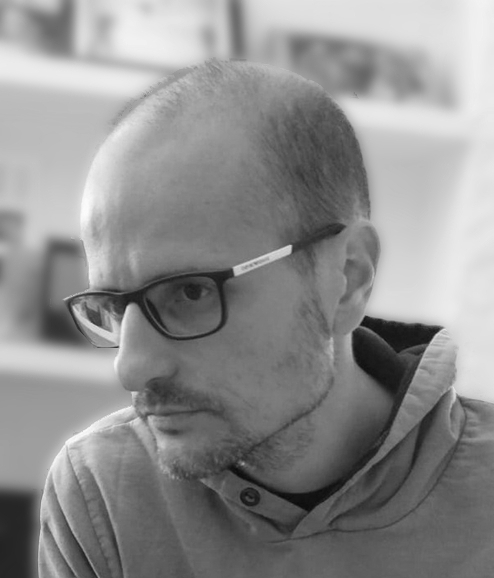}}]{Jes\'us Manj\'on}
is a computer engineer with the UNESCO Chair in Data Privacy and the CRISES
Research Group at the Department of Computer Engineering
and Mathematics at Universitat Rovira i Virgili, Tarragona, Catalonia.
He received his B.Sc. in Computer Engineering in 2004 and his
M.Sc. in Computer Security in 2008. He has participated
in several European and Spanish-funded projects and he is a co-author of
several research publications on security and privacy.
\end{IEEEbiography}


\begin{thebibliography}{99}


\bibitem{BGW} M. Ben-Or, S. Goldwasser and A. Wigderson. 
Completeness theorems for non-cryptographic fault-tolerant
distributed computation. 
In: 20th Annual ACM Symposium on the Theory of Computing - STOC 1988,
pp. 1-10. ACM, 1988.

\bibitem{CCD} D. Chaum, C. Cr\'epeau and I. Damg\aa rd. 
Multiparty unconditionally secure protocols.
In: 20th Annual ACM Symposium on the Theory of Computing - STOC 1988,
pp. 11-19. ACM, 1988.

\bibitem{InfSciRep} J. Domingo-Ferrer, O. Farr\`as, S. Mart\'{\i}nez,
D. S\'anchez and J. Soria-Comas. Self-enforcing protocols via
co-utile reputation management. Information Sciences
367-368 (2016) 159-175.

\bibitem{coutility} J. Domingo-Ferrer, S. Mart\'{\i}nez, D. S\'anchez and 
J. Soria-Comas.
Co-utility: self-enforcing protocols for the mutual benefit
of participants.
Engineering Applications of Artificial Intelligence
59 (2017) 148-158.

\bibitem{evans} D. Evans, V. Kolesnikov and M. Rosulek. 
A Pragmatic Introduction to Secure Multi-Party Computation.
NOW Publishers, 2018.

\bibitem{furukawa} J. Furukawa and Y. Lindell. 
Two-thirds honest-majority MPC for malicious adversaries
at almost the cost of semi-honest. 
In: 26th ACM SIGSAC Conference on Computer and Communications
Security - CCS 2019, pp. 1557-1571. ACM, 2019.

\bibitem{GMW} O. Goldreich, S. Micali and A. Wigderson.
How to play any mental game or a completeness theorem for
protocols with honest majority.
In: 19th Annual ACM Symposium on the Theory of Computing - STOC 1987,
pp. 218-229. ACM, 1987.

\bibitem{harz2019} D. Harz and M. Boman. The scalability of trustless
trust. In: Financial Cryptography and Data Security - FC 2018, pp.
279-193. Springer, 2019.

\bibitem{hastings} M. Hastings, B. Hemenway, D. Noble and
S. Zdancewic. SoK: General purpose compilers for secure
multi-party computation.
In: IEEE Symposium on Security and Privacy - SP 2019, pp. 1220-1237.
IEEE, 2019.


\bibitem{isha06} Y. Ishai, E. Kushilevitz, R. Ostrovsky and A. Sahai.
Cryptography from anonymity. In: 47th Annual IEEE Symposium 
on Foundations of Computer Science - FOCS 2006. IEEE, pp. 239-248.

\bibitem{Kamvar03}
 S. D. Kamvar, M. T. Schlosser and H. Garcia-Molina.
 The EigenTrust algorithm for reputation management in P2P networks.
 In: 12th International Conference on World Wide Web,
pp. 640-651. ACM, 2003.

\bibitem{kupcu15} A. K\"up\c{c}\"u. Incentivized
outsourced computation resistant to malicious contractors.
IEEE Transactions on Dependable and Secure Computing,
14(6) (2015) 633-649.

\bibitem{lindell} Y. Lindell. 
Secure multiparty computation.
{\em Communications of the ACM}, 64(1) (2021) 86-96.

\bibitem{nabi2020} M. Nabi, S. Avizheh, M. V. Kumaramangalam
and R. Safavi-Naini.
Game-theoretic analysis of an incentivized verifiable computation system.
In Financial Cryptography and Data Security -- FC 2019, pp. 50-66.
Springer, 2020.


\bibitem{crowds} M. K. Reiter and A. D. Rubin. Crowds: anonymity for web transactions. 
ACM Transactions on Information and System Security 1 (1998) 66-92.


\bibitem{Y} A. C. Yao. Protocols for secure computations.
In: 23rd Annual Symposium on Foundations of Computer Science - SFCS 1982,
pp. 160-164.
IEEE, 1982.

\end{thebibliography}
\end{document}